\documentclass[a4paper,11pt]{article}
\pdfoutput=1 
\usepackage{jcappub} 

\usepackage[T1]{fontenc} 
\usepackage{verbatim}
\usepackage{csquotes}

\usepackage[style=nature,backend=biber, sorting = none, doi=false, maxbibnames=8, minbibnames=3]{biblatex}

\usepackage[dvipsnames]{xcolor}
\newcommand{\h}{h}

\usepackage[english]{babel}

\usepackage{xcolor}

\usepackage{newtxtext,newtxmath}
\usepackage{ae,aecompl}
\usepackage{enumerate}
\usepackage[switch, modulo]{lineno}
\usepackage[toc,page]{appendix}
\usepackage{soul}

\definecolor{darkgreen}{rgb}{0,0.6,0}
\definecolor{darkpurple}{rgb}{0.4,0,0.6}
\definecolor{darkred}{RGB}{139,0,0}

\usepackage{amssymb}

\usepackage{newtxtext,newtxmath}

\DeclareRobustCommand{\VAN}[3]{#2}
\let\VANthebibliography\thebibliography
\def\thebibliography{\DeclareRobustCommand{\VAN}[3]{##3}\VANthebibliography}

\usepackage{graphicx}	
\usepackage{amsmath}	
\usepackage{float}
\usepackage{orcidlink}
\usepackage{caption}
\usepackage{subcaption}
\usepackage{blindtext}

\title{Detecting Baryon Acoustic Oscillations from Galaxy Clusters: Distance Constraints}

\author[a,b]{Xiaoyun Shao\orcidlink{0009-0002-2863-9351}}
\emailAdd{xiaoyun48@on.br}   
\author[c,d]{Paula S. Ferreira\orcidlink{0000-0002-7540-040X}}
\emailAdd{ferreira\_paula@zju.edu.cn} 
\author[a]{Carlos A. P. Bengaly}
\emailAdd{carlosbengaly@on.br}  
\author[a,g]{Rodrigo S. Gon\c{c}alves}
\emailAdd{rsousa@on.br} 
\author[e,f]{Diego Garcia Lambas}
\emailAdd{diego.garcia.lambas@unc.edu.ar} 
\author[a]{Jailson Alcaniz\orcidlink{0000-0003-2441-1413}}
\emailAdd{alcaniz@on.br}

\affiliation[a]{Observat\'orio Nacional, 20921-400, Rio de Janeiro, RJ, Brasil}
\affiliation[b]{Leinweber Center for Theoretical Physics and Department of Physics,\\
                University of Michigan, 450 Church Street, Ann Arbor, MI 48109, USA}
 \affiliation[c]{Center for Cosmology and Computational Astrophysics, \\ Institute for Advanced Study in Physics, Zhejiang University, Hangzhou 310058, China.}
\affiliation[d]{Institute of Astronomy, School of Physics, Zhejiang University, Hangzhou 310058, China}
\affiliation[e]{Instituto de Astronomia Te\'orica y Experimental (IATE), CONICET-UNC, C\'ordoba, Argentina}
\affiliation[f]{Observatorio Astron\'omico de C\'ordoba (OAC), UNC, C\'ordoba, Argentina}
\affiliation[g]{Departamento de F\'isica, Universidade Federal Rural do Rio de Janeiro, 23897-000, Serop\'edica, RJ,  Brasil}

\abstract{We present a measurement of the baryon acoustic oscillation (BAO) feature using the largest galaxy cluster sample with spectroscopic redshift information to date, comprising 169,679 clusters selected from the WH24 catalogue based on the DESI Legacy Imaging Surveys. We measure the two-point correlation function in two redshift bins, $0.17\leq z<0.35$ and $0.35\leq z<0.5$, corresponding to effective redshifts of $z_{\rm eff}=0.25$ and $0.43$, with the BAO feature detected at approximately the $4\sigma$ level or above
in both bins.
By fitting the measured correlation functions, we obtain dilation parameters of $\alpha=0.97\pm0.07$ and $\alpha=1.01\pm0.04$, corresponding to volume-averaged distance measurements of $D_V/r_d=6.70\pm0.48$ and $D_V/r_d=11.25\pm0.47$, respectively. Using the Planck 2018 sound horizon calibration, we obtain $H_0=66.6 \pm 4.9 \,\mathrm{km\,s^{-1}\,Mpc^{-1}}$ within the flat $\Lambda$CDM framework. We further combine our cluster BAO measurements with cosmic microwave background and Pantheon+ Type Ia supernova observations to constrain dynamical dark energy within the Chevallier--Polarski--Linder  parametrization. The resulting constraints are consistent with $\Lambda$CDM, which lies within the $68\%$ joint credible region in the $w_0$--$w_a$ plane. Our results demonstrate the potential of galaxy clusters as complementary tracers for BAO measurements and cosmological constraints.
}

\begin{document} 
\maketitle
\flushbottom

\section{Introduction}\label{sec:intro}
The discovery of the accelerated expansion of the Universe \cite{riess1998observational,perlmutter1999measurements} has raised fundamental questions about its underlying physical origin. Possible explanations include the presence of an unknown component of dark energy or modifications to general relativity on cosmological scales \cite{Copeland:2006wr,weinberg2013observational}. Distinguishing between these possibilities requires accurate measurements of the cosmic expansion history across a broad range of redshifts. Several observational probes, including the cosmic microwave background (CMB) \cite{aghanim2020planck}, Type Ia supernovae (SNe Ia) \cite{brout2022pantheon+}, and baryon acoustic oscillations (BAO), provide complementary information for investigating the expansion history of the Universe. 
Among these probes, BAO provide a particularly robust measurement of the cosmic expansion history through the large-scale clustering of matter \cite{eisenstein1998baryonic}. 

BAO originate from acoustic waves propagating in the tightly coupled photon--baryon plasma of the early Universe, driven by the interplay between radiation pressure and gravitational attraction. The same underlying acoustic physics gives rise to the characteristic peaks observed in the angular power spectrum of the CMB. These oscillations also imprint a characteristic scale on the late-time matter distribution, determined by the comoving sound horizon at the baryon drag epoch, $r_d \simeq 150\,\mathrm{Mpc}$. This scale is manifested as a prominent peak in the two-point correlation function (2PCF) and a series of oscillatory features in the matter power spectrum. The BAO feature thus provides a standard ruler for measuring cosmological distances as a function of redshift, offering a powerful probe of the expansion history of the Universe \cite{eisenstein1998baryonic}. BAO can also be measured using the angular two-point
correlation function in thin redshift shells, constraining
the transverse distance scale without adopting a fiducial
cosmology to convert redshifts into comoving radial
distances \cite{sanchez2011tracing,carvalho2016baryon,
alcaniz2017measuring,carvalho2020transverse}.
This approach reduces the dependence of the measurement
on the assumed cosmological model, although projection
effects arising from finite shell widths must be assessed.

Since the first detections of BAO in the Sloan Digital Sky Survey (SDSS) and the Two-Degree Field Galaxy Redshift Survey (2dFGRS) \cite{eisenstein2005detection,cole2005two}, BAO measurements have played a central role in observational cosmology. Subsequent large-scale spectroscopic surveys, including WiggleZ \cite{blake2011wigglez}, the Baryon Oscillation Spectroscopic Survey (BOSS) \cite{dawson2013baryon,eisenstein2011sdss}, and its extension eBOSS \cite{dawson2016sdss,blanton2017sloan}, have progressively improved the precision of BAO measurements and extended their redshift coverage. More recently, the Dark Energy Spectroscopic Instrument (DESI) has achieved unprecedented precision in BAO measurements through the clustering of galaxies and quasars, as well as absorption features in the Lyman-$\alpha$ forest \cite{desi2024iii,desi2024iv,desi2025dr2i,desi2025dr2ii}. The BAO measurements from DESI Data Release 1 (DR1) and Data Release 2 (DR2), particularly when combined with CMB and SNe Ia observations, have significantly improved cosmological constraints and provided indications of a possible time evolution in the dark energy equation of state \cite{desi2024vi,desi2025dr2ii}.

Galaxy clusters, the largest gravitationally bound systems in the Universe, provide an alternative tracer of the underlying matter distribution for BAO measurements. As highly biased tracers associated with the highest peaks of the matter density field, galaxy clusters exhibit a stronger clustering amplitude than typical galaxies, making them promising tracers of the BAO feature. BAO measurements from galaxy clusters therefore provide a complementary approach to galaxy-based measurements, allowing the acoustic scale to be investigated using a distinct population of large-scale structure tracers with different clustering properties and observational systematics.

Early studies by Estrada et al. \cite{estrada2009correlation} and H\"utsi \cite{hutsi2010power}, based on the maxBCG cluster catalogue \cite{koester2007maxbcg}, reported tentative detections of the BAO signature in its clustering statistics. Using a spectroscopic sample of 13,904 galaxy clusters at $z \leq 0.4$, Hong et al. \cite{hong2012correlation} detected the BAO feature in the cluster correlation function at a significance of approximately $1.9\sigma$. 
Subsequently, Veropalumbo et al. \cite{veropalumbo2014improved} increased the detection significance to approximately $2.5\sigma$ using 25,226 clusters selected from the catalogue of Wen et al. \cite{wen2012catalog} (hereafter WHL12).  Veropalumbo et al. \cite{veropalumbo2016measuring} further extended this analysis using updated WHL12 samples incorporating spectroscopy from SDSS DR7 \cite{abazajian2009seventh} and DR12 \cite{alam2015eleventh}, achieving BAO detections well above $2\sigma$.
Hong et al. \cite{hong2015detection} also analysed the correlation function of 79,091 galaxy clusters selected from SDSS photometric catalogues \cite{wen2012catalog,wen2015calibration} (hereafter WH15), supplemented with SDSS-III DR12 spectroscopic data \cite{alam2015eleventh}, obtaining a BAO detection significance of $3.7\sigma$ at $z\leq0.5$.

In this work, we investigate the BAO signal in the 2PCF of 169,679 galaxy clusters selected from the Wen \& Han (2024, hereafter WH24) catalogue \cite{wen2024catalog}, constituting the largest spectroscopic sample of galaxy clusters to date. The catalogue was constructed from the DESI Legacy Imaging Surveys DR9 and DR10 \cite{dey2019overview}, with spectroscopic redshift information supplemented by the Two Micron All-Sky Survey (2MASS) Redshift Survey \cite{huchra20122mass}, SDSS DR17 \cite{abdurro2022seventeenth}, and the early data release of DESI \cite{desi2024early}.
We measure the cluster two-point correlation function,
$\xi(s)$, to obtain isotropic BAO distance measurements,
$D_V/r_d$, and subsequently derive constraints on
cosmological parameters.

The remainder of this paper is organized as follows. Section~\ref{sec:data} describes the galaxy cluster samples employed in our analysis. In Section~\ref{sec:methods}, we outline the methodology adopted to estimate the two-point correlation functions and model the BAO signal. Section~\ref{sec:results} presents our clustering and BAO measurements, together with the resulting cosmological constraints. Finally, Section~\ref{sec:conclusions} summarizes our main findings and conclusions.

\section{Data}\label{sec:data}
In this section, we describe the galaxy cluster samples used in our clustering analysis. For further details on the cluster identification and the construction of the cluster samples, we refer the reader to Wen \& Han (2024) \cite{wen2024catalog}.

The WH24 cluster catalogue contains 1,581,179 clusters, including 877,806 newly identified systems, covering a total sky area of approximately \(24,000~{\rm deg}^2\) and extending to redshifts of \(z\sim1.5\), making it the largest cluster catalogue currently available. The WH24 catalogue also exhibits high completeness and purity for low-redshift and rich clusters, with a detection rate exceeding 90\% for clusters with masses above \(10^{14}\,M_\odot\) when cross-matched with existing catalogues \cite{wen2024catalog,yuan2025sample}.
The cluster identification is based on the stellar-mass distribution of galaxies within redshift slices around preselected massive galaxies. Structures are identified starting from the positions of the observed galaxies in the photometric samples of the DESI Legacy Imaging Surveys DR9 and DR10~\cite{dey2019overview}. A galaxy cluster is identified if its richness is \(\lambda_{500}\geq10\) and at least six member-galaxy candidates are found within \(r_{500}\), the radius within which the mean enclosed density is 500 times the critical density of the Universe. 
The cluster centre is defined by the angular position of the brightest cluster galaxy (BCG), while the photometric redshift is estimated by averaging the redshifts of the member galaxies. 
A cluster is assigned a spectroscopic redshift when a spectroscopic measurement is available for its BCG. The catalogue contains 338,841 clusters with spectroscopic redshifts, providing a large spectroscopic sample for clustering analyses.

An accurate determination of cluster redshifts is essential for measurements of the large-scale spatial distribution of galaxy clusters. The larger uncertainties associated with photometric redshifts introduce substantial distortions along the line of sight and consequently affect measurements of 2PCF, complicating their cosmological interpretation \cite{marulli2012cosmology,sereno2015new}. We therefore restrict our clustering analysis to clusters with spectroscopic redshift information.

The spectroscopic redshifts adopted in WH24 are compiled from the 2MASS Redshift Survey \cite{huchra20122mass}, SDSS DR17 \cite{abdurro2022seventeenth}, and the early data release of DESI \cite{desi2024early}. Together, these surveys provide spectroscopic redshifts for approximately 3.1 million galaxies in the DESI Legacy Imaging Surveys, yielding accurate redshift information for a large number of cluster members. Since the cluster centre is defined by the position of the BCG, no distinction is made hereafter between the angular position of a cluster and that of its BCG. 
Regions affected by poor or failed spectroscopic observations are excluded using the corresponding angular masks.

At very low redshift, peculiar velocities introduce additional uncertainty in the inferred radial distances. We therefore impose a lower redshift limit of \(z\geq0.17\), which also excludes the regime dominated by the 2MASS Redshift Survey. Since most of the cluster spectroscopic redshifts are provided by SDSS DR17, with a smaller contribution from the DESI early data release, we further restrict the sample to the SDSS spectroscopic footprint.
We adopt an upper redshift limit of (z$\leq$0.5) as a conservative choice regarding sample completeness  and to facilitate comparison with Hong et al. \cite{hong2015detection}, who measured the BAO feature using 79,091 clusters at (z$\leq$0.5) from the WH15 catalogue with SDSS spectroscopic redshifts. The resulting sample used in the present paper contains 169,679 clusters, more than twice the number used in their analysis.

\begin{figure}
    \centering
    \begin{minipage}[c]{0.48\linewidth}
        \centering
        \includegraphics[width=\linewidth]{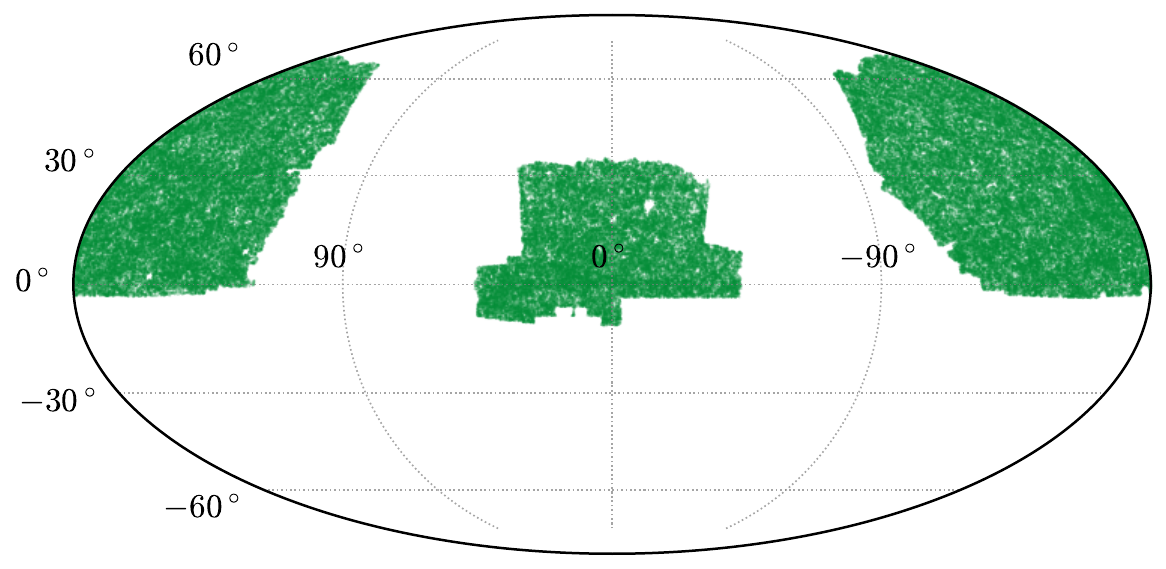}
    \end{minipage}
    \hfill
    \begin{minipage}[c]{0.48\linewidth}
        \centering
        \includegraphics[width=\linewidth]{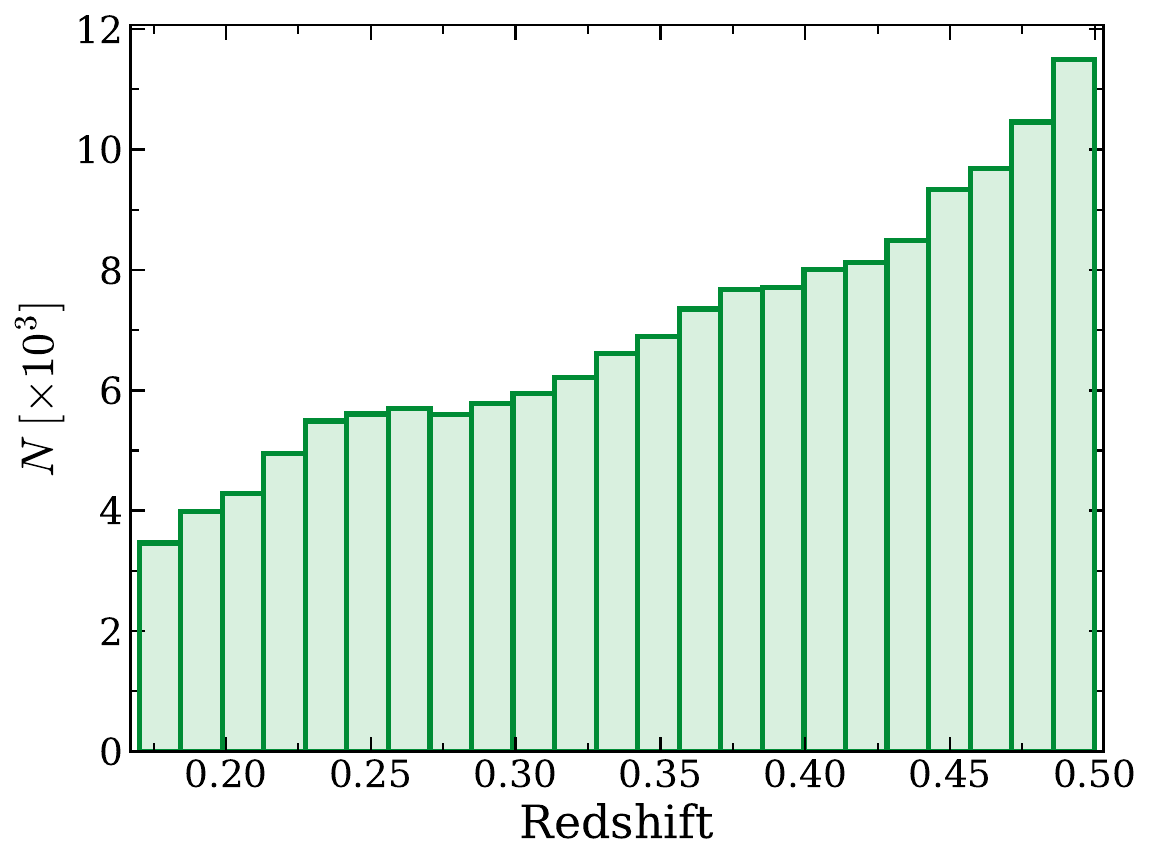}
    \end{minipage}
    \caption{
    Sky distribution and redshift distribution of the cluster sample.
    The left panel shows the distribution of clusters in right ascension 
    and declination, while the right panel shows the corresponding
    redshift distribution. The redshift range is restricted to
    $0.17<z<0.50$. 
    }
    \label{fig:nz}
\end{figure}

The sky coverage and redshift distribution of the full cluster sample are shown in Figure~\ref{fig:nz}.
We  further divide the catalogue into two redshift subsamples, \(0.17\leq z<0.35\) and \(0.35\leq z<0.5\), as summarized in Table~\ref{tab:cluster_samples}. The effective volumes are calculated following the prescription of Reid et al. \cite{reid2016sdss}, yielding \(V_{\rm eff}=0.66\) and \(0.97\,h^{-3}\,{\rm Gpc}^3\) for the low- and high-redshift subsamples, respectively. 
We also account for the mass dependence of cluster and sample completeness by assigning each cluster a combined weight based on a mass weight and a completeness weight, following the approach of Hong et al. \cite{hong2015detection}. In addition, we apply the Feldman--Kaiser--Peacock (FKP) weight ~\cite{feldman1993power} to account for variations in the tracer number density with redshift,  following Reid et al. \cite{reid2016sdss}. We assess the impact of these weights by comparing the clustering measurements with and without weighting, finding no significant difference in the inferred BAO scale.

\begin{table*}
\centering
\caption{
The main properties of the cluster samples used for the clustering analysis. The bias is obtained by modelling the measured 2PCF over the separation range $10<r_p<30      \h^{-1}\mathrm{Mpc}$ (see Section~\ref{sec:bias_fit}).
}
\label{tab:cluster_samples}
\begin{tabular}{ccccc}
\hline
Redshift range & Effective redshift & Number of clusters & $b$ & $V_{\rm eff}\,[h^{-3}\,\mathrm{Gpc}^3]$ \\
\hline

$0.17 \le z < 0.35$ & 0.25 & 78,232 & $2.51 \pm 0.03$ & 0.66 \\
$0.35 \le z \le 0.5$ & 0.43 & 91,447 & $2.77 \pm 0.03$ & 0.97 \\
\hline
\end{tabular}
\end{table*}

\section{Method}\label{sec:methods}
In this section, we present the methodology for measuring and modelling the clustering of galaxy clusters. We estimate the redshift-space 2PCF, determine the cluster bias, and model the BAO feature to constrain the dilation parameter and cosmological distance scale.

\subsection{The two-point correlation function}\label{sec:2pcf}

We characterize the clustering of the cluster samples using the redshift-space two-point correlation function, $\xi(s)$, estimated with the Landy--Szalay estimator \cite{landy1993bias},

\begin{equation}
\xi(s)=\frac{DD(s)-2DR(s)+RR(s)}{RR(s)},
\label{eq:ls_ksi}
\end{equation}
where $DD(s)$, $DR(s)$, and $RR(s)$ denote the weighted numbers of data--data, data--random, and random--random pairs, respectively, within a separation bin of $s \pm \Delta s/2$, where $\Delta s=5\,h^{-1}\mathrm{Mpc}$.   The quantities $N_{DD}$, $N_{DR}$, and $N_{RR}$ are the corresponding weighted normalization factors. The random catalogue is constructed to be 16 times larger than the data catalogue, thereby reducing the impact of shot noise in the pair-count measurements. It covers the same sky area and follows the same redshift distribution as the observed cluster sample. We measure the correlation function and its uncertainty in 32 separation bins spanning $40$--$200 h^{-1} \mathrm{Mpc}$. \footnote{We also perform the analysis using a coarser binning of 16 separation bins over the same range and find no significant changes in the resulting BAO distance measurements.}

To convert the observed redshifts into comoving distances and determine the three-dimensional separations between cluster pairs, we adopt a spatially flat $\Lambda$CDM fiducial cosmology based on the Planck 2018 results \cite{aghanim2020planck}, where we fix the cosmological parameters to $\Omega_{\rm b}=0.0491$, $n_{\rm s}=0.9649$, $\sigma_8=0.811$, $\Omega_{\rm m}=0.3121$, $\Omega_{\Lambda}=0.6879$, and $H_0=67.51 \,\mathrm{km\,s^{-1}\,Mpc^{-1}}$.

\subsection{Cluster bias }\label{sec:bias}
The bias of the cluster samples is inferred from the projected two-point correlation function. Projecting the correlation function along the line-of-sight direction substantially suppresses the effects of redshift-space distortions. We model the projected correlation function as
\begin{equation}
w_p(r_p)=b^2\int_{0}^{\pi_{\rm max}}
\xi^{\rm DM}(r_p,\pi)\,{\rm d}\pi,
\label{eq2}
\end{equation}
where $\pi_{\rm max}$ is chosen to be the same as that adopted in the measurement of $w_p(r_p)$, and $\xi^{\rm DM}(r_p,\pi)$ denotes the redshift-space dark-matter correlation function as a function of the separations perpendicular and parallel to the line of sight. Redshift-space distortions are modelled using the dispersion model \cite{kaiser1987clustering,hamilton1992measuring,davis1983survey,marulli2012clustering}, and the  dark-matter correlation function is obtained by Fourier transforming the matter power spectrum computed with CAMB \cite{lewis2000efficient}.

\subsection{Covariance matrix}

The covariance matrix of the correlation function is estimated using log-normal mock catalogues \cite{coles1991lognormal}. This approach has been widely adopted in clustering and BAO analyses \cite{beutler20116df,blake2011wigglez,chuang2015nifty}.
The density-field realizations are generated using the redshift-space  power-spectrum model
\begin{equation}
P_{\rm model}(k)
=
b^2
\left(
1+\frac{2}{3}\beta+\frac{1}{5}\beta^2
\right)
P_{\rm DM}(k),
\label{eq:pmodel}
\end{equation}
where $P_{\rm DM}(k)$ is the  matter power spectrum computed with the CAMB \cite{lewis2000efficient}, $b$ is the linear bias parameter, and $\beta=f/b$, with the linear growth rate approximated as $f=\Omega_{\rm M}(z)^{0.545}$ \cite{linder2007parameterized}. 
The bias parameter is determined from the projected cluster correlation function, following the procedure described in Sec.~\ref{sec:bias}, with the covariance matrix estimated using the jackknife method. The resulting bias measurements and their corresponding uncertainties for each sample are reported in Table~\ref{tab:cluster_samples}.

The log-normal mock catalogues are generated in simulation boxes sufficiently large to cover the entire survey volume of the cluster sample, using a regular grid with a cell size equal to half the bin width used for the 2PCF measurements. The survey selection function is incorporated into the mock generation to reproduce the angular and radial selection of the observed cluster sample. The resulting log-normal density fields are smooth on scales smaller than the cell size. In total, 1000 mock cluster catalogues are generated, and the 2PCF is measured independently for each realization.

The covariance matrix is estimated from the mock catalogues as

\begin{equation}
C_{ij}
=
\frac{1}{N-1}
\sum_{k=1}^{N}
\left(\xi_i^{k}-\bar{\xi}_i\right)
\left(\xi_j^{k}-\bar{\xi}_j\right),
\label{eq:covariance}
\end{equation}
where $N=1000$ is the number of mock catalogues, $\xi_i^{k}$ is the correlation-function measurement in the $i$th separation bin for the $k$th mock realization, and $\bar{\xi}_i$ is the mean correlation function over all $N$ mock catalogues in the same bin.

\subsection{Model}
To extract cosmological constraints from the BAO peak, the correlation function is modelled following \cite{anderson2014clustering} as
\begin{equation}
\xi_{\rm model}(r)
=
B^2 \xi_{\rm DM}(\alpha r)
+
A_0
+
\frac{A_1}{r}
+
\frac{A_2}{r^2},
\label{eq:xi_model}
\end{equation}
where $\xi_{\rm DM}(r)$ denotes the dark-matter correlation-function template evaluated for the fiducial cosmology. The parameter $B$ sets the overall clustering amplitude. Since the comoving separations are inferred assuming a fiducial cosmology, a mismatch between the fiducial and true cosmologies leads to a geometrical rescaling of the measured BAO scale. This effect is captured by the dilation parameter $\alpha$, which rescales the separation in the template and therefore quantifies the shift of the BAO peak relative to that expected in the fiducial cosmology. The coefficients $A_0$, $A_1$, and $A_2$ describe a smooth broadband contribution, allowing residual scale-dependent effects that are not captured by the BAO template to be marginalized over.

We model the template power spectrum, accounting for the nonlinear broadening of the BAO feature, as
\begin{equation}
P_{\rm DM}(k)
=
\left[P_{\rm lin}(k)-P_{\rm nw}(k)\right]
\exp\left(-\frac{k^2\Sigma_{\rm NL}^2}{2}\right)
+
P_{\rm nw}(k),
\label{eq:pdw}
\end{equation}
where $P_{\rm lin}(k)$ is the linear matter power spectrum computed with CAMB \cite{lewis2000efficient}, and $P_{\rm nw}(k)$ is the corresponding smooth, no-wiggle spectrum obtained using the prescription of Eisenstein \& Hu \cite{eisenstein1998baryonic}. The parameter $\Sigma_{\rm NL}$ accounts for the nonlinear broadening of the BAO feature and is allowed to vary in the fit.
The corresponding correlation function template is obtained by Fourier transforming the matter power spectrum:
\begin{equation}
\xi_{\rm DM}(r)
=
\frac{1}{2\pi^2}
\int dk\,k^2 P_{\rm DM}(k)
\frac{\sin(kr)}{kr}.
\label{eq:xi_dm}
\end{equation}
The posterior distributions of the model parameters are constrained using Markov Chain Monte Carlo (MCMC) sampling assuming a Gaussian likelihood, with

\begin{equation}
\chi^2
=
\sum_{i,j}
\left[\xi_i-\xi_{\rm model}(r_i)\right]
C^{-1}_{ij}
\left[\xi_j-\xi_{\rm model}(r_j)\right].
\label{eq:chi2}
\end{equation}
where $\xi_i$ is the measured 2PCF in the $i$th separation bin and $C_{ij}$ is the covariance matrix.

\begin{figure}
    \centering
    \includegraphics[width=0.48\textwidth]{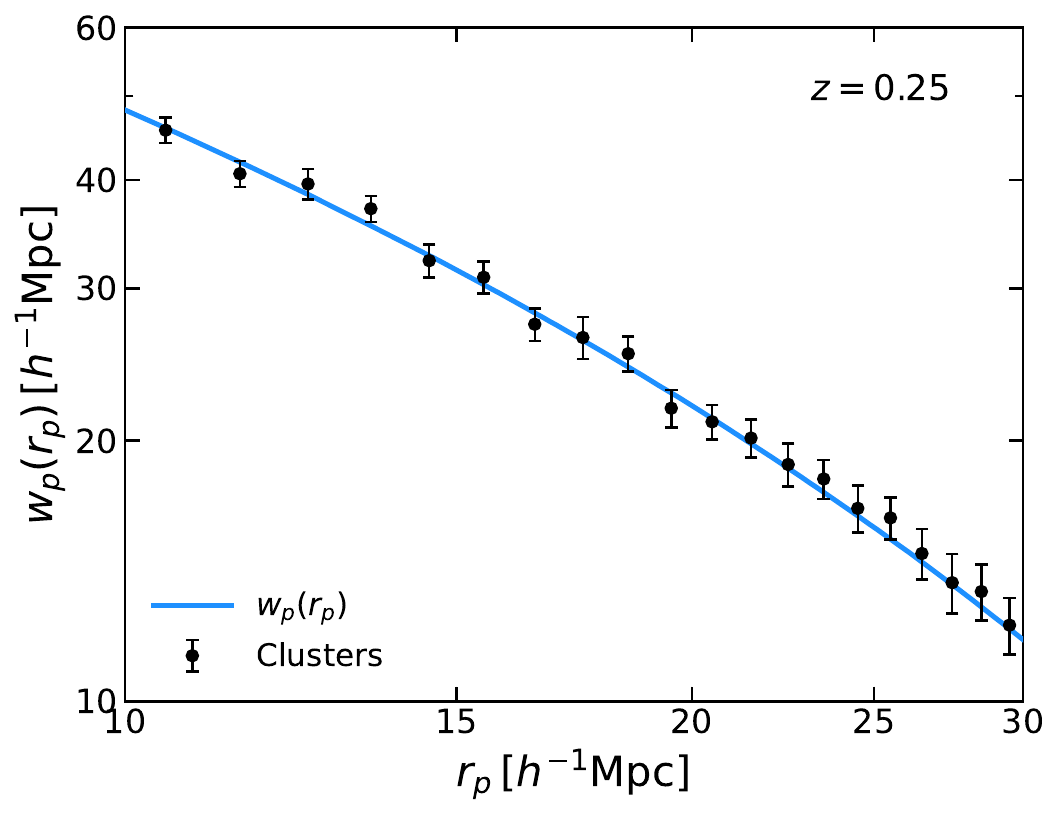}
    \includegraphics[width=0.48\textwidth]{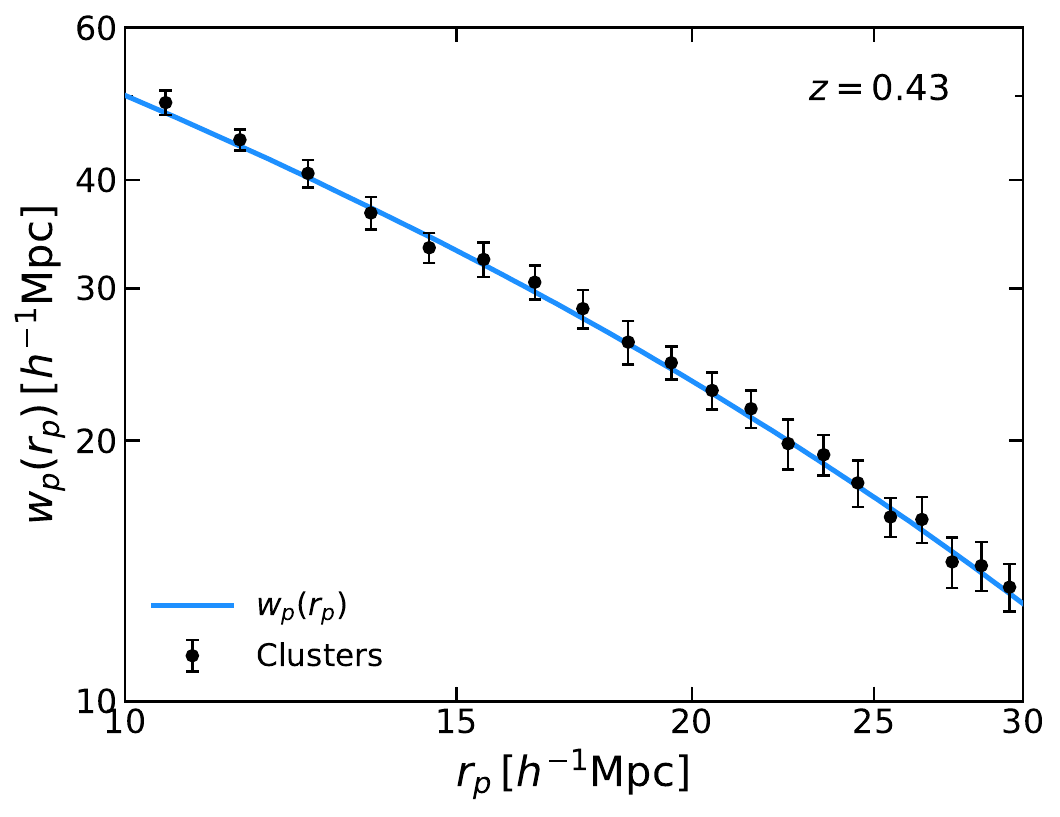}
\caption{Projected two-point correlation functions of the two cluster samples at $z_{\rm eff}=0.25$ (left) and $z_{\rm eff}=0.43$ (right). Black points with error bars show the measurements, while the blue solid curves represent the corresponding best-fitting models.}
    \label{fig:wp}
\end{figure}

\section{Results}\label{sec:results}
In this section, we present the main results of our analysis. We first measure the projected correlation function and constrain the linear bias of the cluster samples. We then investigate the BAO signal, derive cosmological distance constraints, and compare our measurements with previous studies. Finally, we combine our cluster BAO measurements with CMB and SNe Ia data to constrain dynamical dark energy.

\subsection{Projected correlation function and bias}\label{sec:bias_fit}

Figure~\ref{fig:wp} presents the projected correlation functions, $w_p(r_p)$, for the two cluster samples. The measurements are shown as black points with error bars, while the blue solid curves correspond to the best-fitting theoretical models. The linear bias is inferred following the procedure described in Section~\ref{sec:bias}, with the fits restricted to the range $10<r_p<30    h^{-1}\mathrm{Mpc}$. In our analysis, the integration is performed up to $\pi_{\rm max}=50 h^{-1}\mathrm{Mpc}$. We also test different values of $\pi_{\rm max}$ and find that the resulting variations are negligible compared with the statistical uncertainties.

As shown in Table~\ref{tab:cluster_samples}, we obtain $b=2.51\pm0.03$ with $\chi^2_{\rm min}/{\rm dof}=25.94/16$ for the low-redshift sample at $z_{\rm eff}=0.25$, and $b=2.77\pm0.03$ with $\chi^2_{\rm min}/{\rm dof}=15.64/16$ for the high-redshift sample at $z_{\rm eff}=0.43$. The inferred bias values are higher than those typically measured for galaxy samples, as expected for galaxy clusters, which preferentially trace the highest-density peaks of the underlying matter distribution rather than representing a subsample of galaxies \cite{veropalumbo2014improved}.
The best-fitting bias values are subsequently adopted when generating the lognormal mock catalogues used to estimate the covariance matrix of the measured two-point correlation function.

\subsection{Clustering measurements }\label{sec:mea}
\begin{figure}
    \centering
    \includegraphics[width=0.48\textwidth]{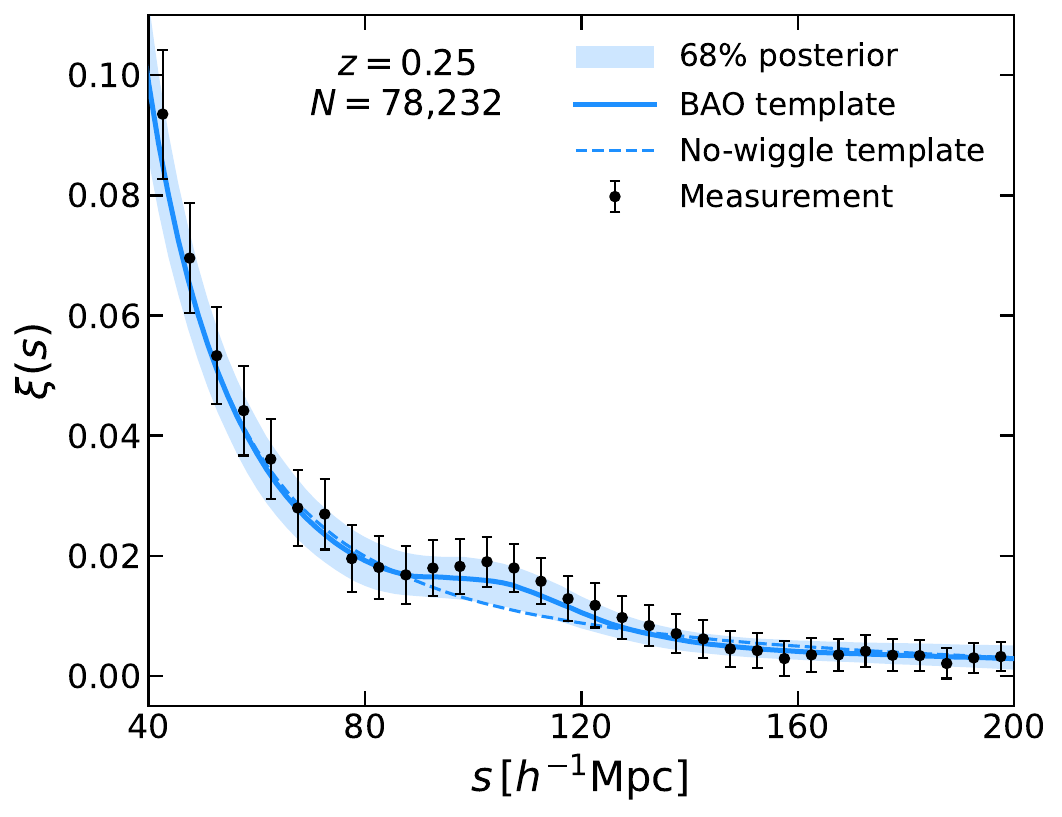}
    \includegraphics[width=0.48\textwidth]{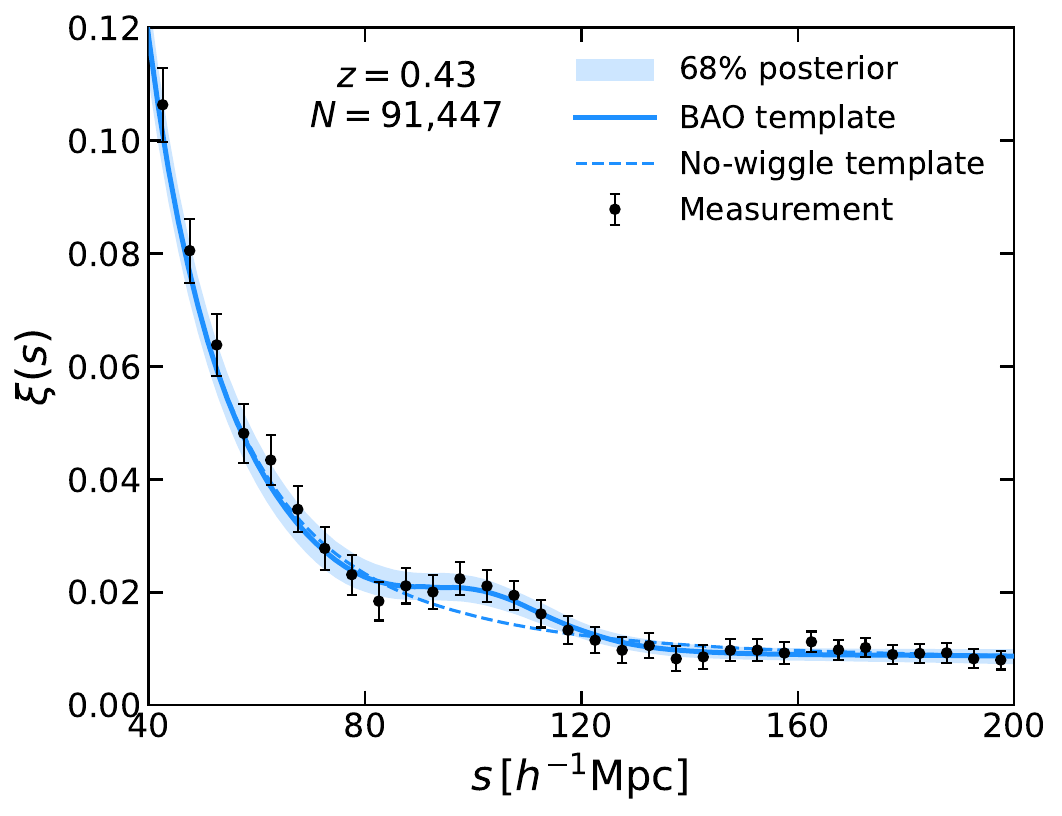}
    \caption{
    Redshift-space 2PCF for the cluster samples at $z=0.25$ (left) and $z=0.43$ (right). The black data points show the measured correlation functions, with uncertainties estimated from lognormal mock catalogues. The solid blue lines indicate the best-fit BAO templates, while the dashed blue lines show the corresponding no-wiggle templates. The shaded regions represent the 68\% posterior credible intervals.
    }
    \label{fig:ksi}
\end{figure}

We present the measured redshift-space 2PCF  for the two cluster samples considered, at effective redshifts  $z=0.25$ (left) and $z=0.43$ (right) in Fig.~\ref{fig:ksi}. The black data points show the measured 2PCF, with error bars estimated from lognormal mock catalogues. The solid blue curves represent the best-fit BAO templates, while the dashed blue curves show the corresponding no-wiggle models.
The shaded regions indicate the 68\% posterior credible intervals.

Despite the relative sparseness of the cluster samples, the clustering
signal is well measured in both redshift bins. Both 2PCFs exhibit a
BAO feature at separations around $100\,h^{-1}\mathrm{Mpc}$ and are
well described by the best-fit BAO templates.

Previous studies have investigated the reliability of jackknife covariance estimates in clustering analyses. By comparing covariance matrices obtained with internal and external error-estimation methods over scales of $0.1$--$40 h^{-1}\mathrm{Mpc}$, Norberg et al. \cite{norberg2009statistical} found that the jackknife method tends to overestimate the variance on small scales, $\lesssim 2$--$3  h^{-1}\mathrm{Mpc}$, although it provides more reliable estimates on scales $\gtrsim 10  h^{-1}\mathrm{Mpc}$. At the much larger scales relevant for BAO measurements, $\sim100 h^{-1}\mathrm{Mpc}$, Beutler et al. \cite{beutler20136df} showed that, for the 6dFGS galaxy sample, jackknife errors are noisier and generally larger than those derived from lognormal mocks.

These findings have also been confirmed in analyses based specifically on cluster samples. Veropalumbo et al. \cite{veropalumbo2016measuring} found that covariance matrices obtained from lognormal mocks are considerably smoother than their jackknife counterparts, particularly when a large number of realizations is employed. Similarly, Hong et al. \cite{hong2015detection} compared covariance matrices estimated from jackknife resampling and lognormal mocks using galaxy clusters with SDSS-III spectroscopic redshifts. They found that jackknife uncertainties are larger in most separation bins and that the jackknife covariance matrices are considerably noisier.

\begin{figure}
    \centering
\includegraphics[width=0.45\textwidth]{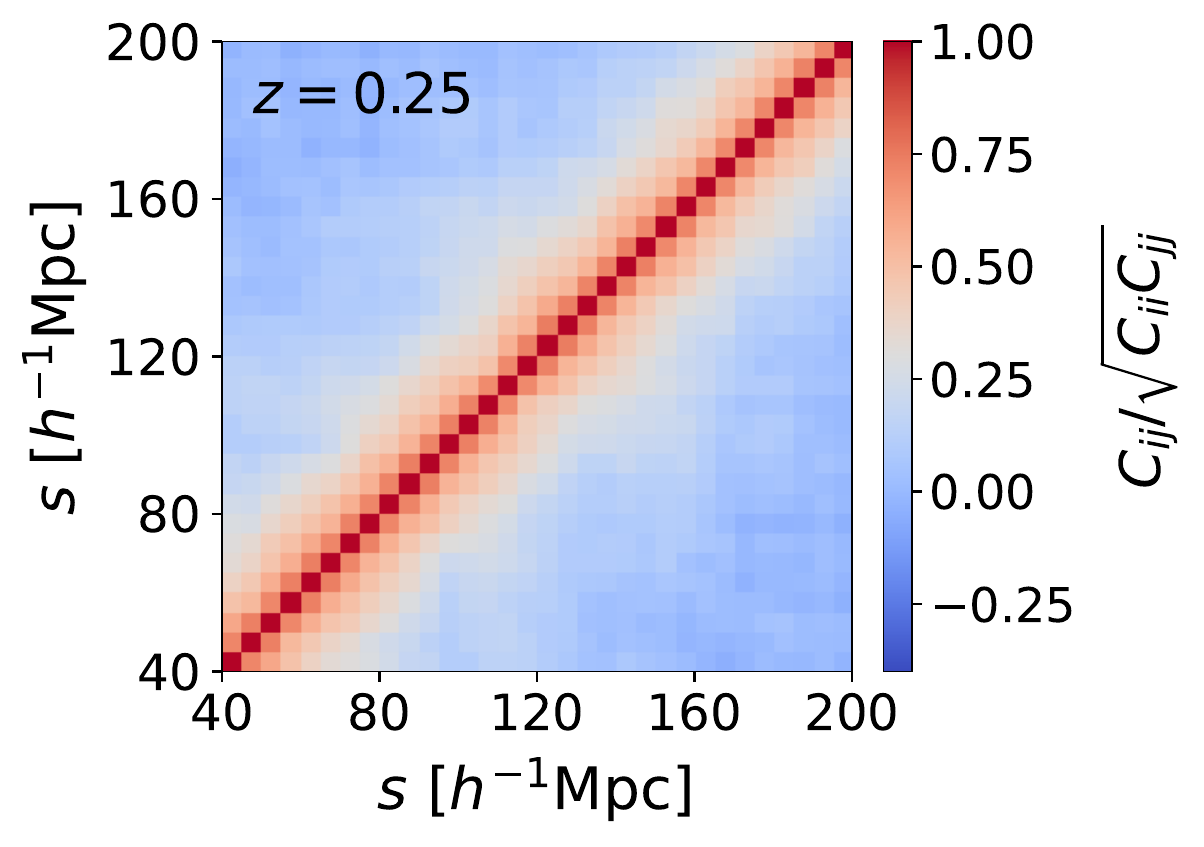}
\includegraphics[width=0.45\textwidth]{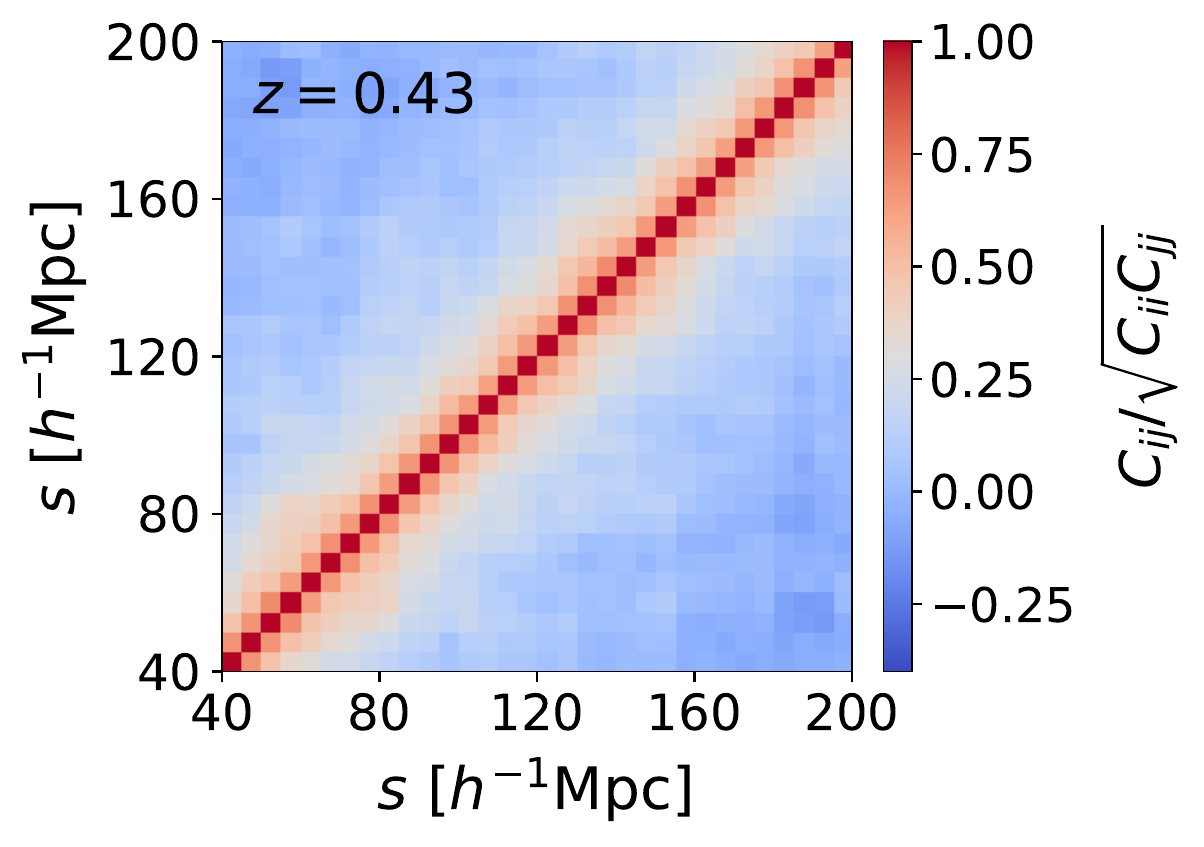}
    \caption{
Correlation matrices of the redshift-space two-point correlation function, $\xi(s)$, for the cluster samples at $z_{\rm eff}=0.25$ (left) and $z_{\rm eff}=0.43$ (right). The correlation coefficients are defined as $r_{ij}=C_{ij}/\sqrt{C_{ii}C_{jj}}$, where $C_{ij}$ is the covariance between the $i$th and $j$th separation bins. The covariance matrices are estimated from the lognormal mock catalogues. 
    }
    \label{fig:cov}
\end{figure}

Given that our analysis focuses on the BAO regime and involves a relatively sparse cluster sample, we therefore adopt the lognormal-mock approach to obtain a more stable covariance estimate. Specifically, we use 1000 independent lognormal mock realizations to estimate the covariance matrix and the corresponding statistical uncertainties, as shown in Fig.~\ref{fig:cov}.

\subsection{Isotropic BAO constraints}
We constrain the BAO distance scale by fitting the position of the BAO feature in the measured 2PCF over the separation range $40<s<200\,h^{-1}\mathrm{Mpc}$  in the two redshift intervals. \footnote{We also test a narrower fitting range of $50<s<180\,h^{-1}\mathrm{Mpc}$ and find no significant changes in the inferred BAO dilation parameters $\alpha$.} A clear BAO signal is detected in all cluster samples, and the corresponding fitting results are summarized in Table~\ref{tab:alpha}. We obtain BAO dilation parameters of $\alpha=0.97\pm0.07$ and $\alpha=1.01\pm0.04$ at redshifts of $z=0.25$ and $z=0.43$, respectively, both consistent with unity within the uncertainties.

\begin{figure}
    \centering
    \includegraphics[width=0.7 \linewidth]{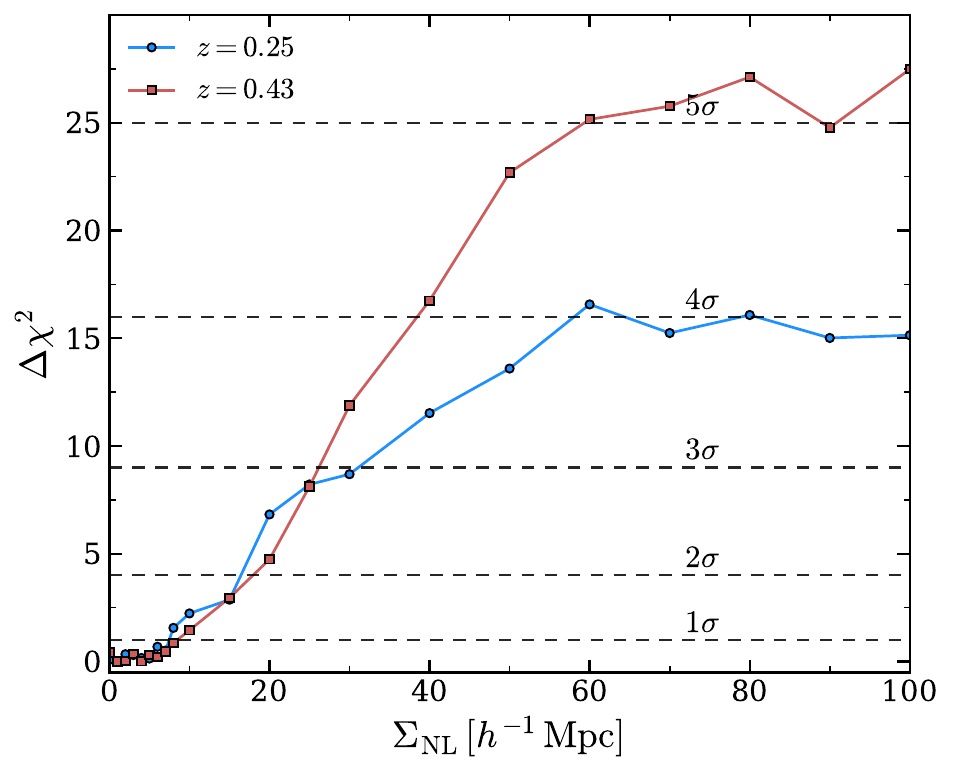}
    \caption{
    Values of $\Delta\chi^2$ as a function of the nonlinear damping scale $\Sigma_{\rm NL}$ for the two cluster samples at $z=0.25$ (blue circles) and $z=0.43$ (red squares). The $\Delta\chi^2$ values are measured relative to the minimum $\chi^2$ for each sample and span from the linear-power-spectrum limit, $\Sigma_{\rm NL}=0~h^{-1}\mathrm{Mpc}$, toward increasingly damped BAO models at large $\Sigma_{\rm NL}$. The horizontal dashed lines indicate significance levels.}
    \label{fig:signal_nl}
\end{figure}

The model includes the nonlinear damping parameter $\Sigma_{\rm NL}$, which characterizes the suppression of the BAO feature in the power spectrum due to nonlinear evolution. To assess the sensitivity of the BAO detection to this parameter, we refit the measured 2PCF while varying $\Sigma_{\rm NL}$ from $0$ to $100~h^{-1}\mathrm{Mpc}$. Figure~\ref{fig:signal_nl} shows the resulting $\Delta\chi^2$ as a function of $\Sigma_{\rm NL}$, where each point is defined relative to the absolute minimum $\chi^2$ of the corresponding curve. Since sufficiently large values of $\Sigma_{\rm NL}$ strongly damp the oscillatory feature and effectively approach a no-BAO model, the increase in $\chi^2$ provides a direct indication of the significance of the BAO detection.

We detect the BAO feature at approximately the $4\sigma$ level
or above in both redshift bins. This exceeds the previous $3.7\sigma$ detection reported by Hong et al. \cite{hong2015detection}, based on 79,091 galaxy clusters selected from the WH15 catalogue over the full redshift range $z\leq0.5$, compared with the 169,679 clusters used in our analysis.

At the same time, models with $\Sigma_{\rm NL}\lesssim 8\,h^{-1}\mathrm{Mpc}$ cannot be clearly distinguished, as they are all consistent within $1\sigma$. Nevertheless, Fig.~\ref{fig:signal_nl} indicates that the preferred nonlinear damping is relatively weak, with the minimum $\chi^2$ occurring in some cases close to $\Sigma_{\rm NL}=0\,h^{-1}\mathrm{Mpc}$. This behavior is consistent with previous studies suggesting that galaxy clusters, as highly biased tracers of the matter distribution, experience less nonlinear degradation of the BAO feature than galaxy samples, implying that density-field reconstruction is expected to provide only limited improvements in BAO distance measurements from galaxy clusters \cite{marulli2017redshift,veropalumbo2014improved,veropalumbo2016measuring}.\footnote{Veropalumbo (2016, Fig.~3.4) \cite{veropalumbo2016clustering} also showed that the BAO peak position remains approximately unchanged after density-field reconstruction, with no significant improvement in the BAO detection significance for galaxy clusters.} In the following analysis, we adopt the lognormal-mock results with $\Sigma_{\rm NL}=4\,h^{-1}\mathrm{Mpc}$ as our fiducial distance constraints.

\begin{table}
\centering
\caption{
BAO fitting results for the cluster samples. The columns list the redshift range, effective redshift, BAO dilation parameter $\alpha$, and isotropic BAO distance $D_V/r_d$. The constraints are obtained by fitting the measured 2PCF over the separation range $40<s<200 \h^{-1}\mathrm{Mpc}$, with the nonlinear damping scale fixed at $\Sigma_{\rm NL}=4 h^{-1}\mathrm{Mpc}$. 
}
\label{tab:alpha}
\begin{tabular}{ccc}
\hline
Effective redshift & $\alpha$ & $D_V/r_d$ \\
\hline
0.25 & $0.97 \pm 0.07$ & $6.70 \pm 0.48$ \\
0.43 & $1.01 \pm 0.04$ & $11.25 \pm 0.47$ \\
\hline
\end{tabular}
\end{table}

The fitted dilation parameter $\alpha$ can be translated into an isotropic BAO distance measurement by comparing the observed BAO scale with that of the fiducial cosmology. The volume-averaged distance at the effective redshift is related to $\alpha$ through
\begin{equation}
D_V(\bar{z})
=
\alpha D_V^{\rm fid}(\bar{z})
\frac{r_d}{r_d^{\rm fid}},
\label{eq:dv_alpha}
\end{equation}
where $r_d$ and $r_d^{\rm fid}$ denote the sound horizon at the baryon drag epoch in the true and fiducial cosmologies, respectively. The isotropic volume-averaged distance is defined as
\begin{equation}
D_V(z)
=
\left[
(1+z)^2 D_A^2(z)
\frac{cz}{H(z)}
\right]^{1/3},
\label{eq:dv}
\end{equation}
where $c$ is the speed of light, $D_A(z)$ is the angular diameter distance, and $H(z)$ is the Hubble parameter.

\begin{figure}
    \centering
    \includegraphics[width=0.7\linewidth]{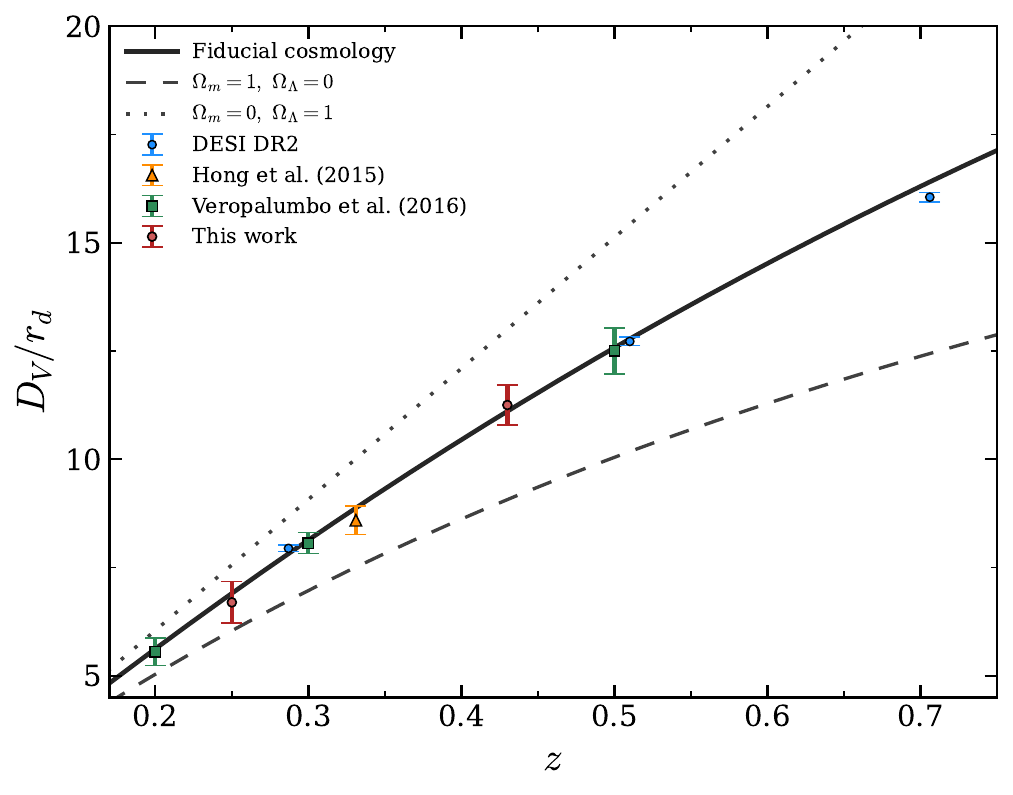}
    \caption{
Isotropic BAO distance measurements, $D_V/r_d$, as a function of redshift. The red points show the cluster BAO measurements obtained in this work, while the orange and green points represent previous cluster BAO measurements from Hong et al.~(2015) \cite{hong2015detection} and Veropalumbo et al.~(2016) \cite{veropalumbo2016measuring}, respectively. The blue points correspond to the DESI DR2 measurements \cite{abdul2025desi}, shown only up to $z=0.7$ to focus on the redshift range relevant to cluster BAO measurements. The DESI DR2 point near $z=0.3$ is shifted horizontally by $\Delta z=-0.008$ for visual clarity. The solid black line shows the prediction of the fiducial cosmological model, while the dashed and dotted lines correspond to $(\Omega_m,\Omega_\Lambda)=(1,0)$ and $(0,1)$, respectively.
    }
    \label{fig:dv_rd}
\end{figure}

Using the fitted values of $\alpha$ and Eq.~\ref{eq:dv_alpha}, we obtain $D_V/r_d=6.70\pm0.48$ and $11.25\pm0.47$ at $z=0.25$ and $z=0.43$, respectively, as listed in Table~\ref{tab:alpha}.
Figure~\ref{fig:dv_rd} compares our measurements (red points) with previous cluster BAO results from Hong et al.~(2015) \cite{hong2015detection} (orange points) and Veropalumbo et al.~(2016) \cite{veropalumbo2016measuring} (green points), as well as the DESI DR2 galaxy BAO measurements \cite{abdul2025desi} (blue points). The solid black curve shows the theoretical prediction for the Planck 2018 $\Lambda$CDM cosmology \cite{aghanim2020planck}. Our cluster-based distance measurements are consistent with previous cluster BAO results, the DESI DR2 measurements at comparable redshifts, and the distance--redshift relation predicted by the standard $\Lambda$CDM cosmology.

\subsection{Cosmological constraints from cluster BAO}
\label{sec:cosmo_constraints}

The distance measurements derived above can be used to constrain cosmological parameters through their dependence on $D_V(z)$, as given in Eq.~\ref{eq:dv}. Given that our analysis relies on only two BAO distance measurements, we do not expect the resulting constraints to be competitive with those derived from larger galaxy surveys or combinations of multiple cosmological probes. Instead, the aim of this analysis is to assess the consistency of our cluster BAO measurements with the predictions of the standard $\Lambda$CDM cosmology.

We calibrate the BAO distance measurements using the Planck 2018 constraint on the sound horizon at the baryon drag epoch, $r_d=147.09\pm0.26\,\mathrm{Mpc}$ \cite{aghanim2021erratum}, obtaining $D_V(0.25)=984.81\pm70.43\,\mathrm{Mpc}$ and $D_V(0.43)=1654.91\pm68.64\,\mathrm{Mpc}$. We consider a spatially flat $\Lambda$CDM cosmology, characterized by a dark-energy equation of state $w=-1$ and a negligible contribution from radiation over the redshift range relevant to our measurements. In this model, the Hubble expansion rate is given by ${H^2(z)}
={H_0^2}[\Omega_{\rm m}(1+z)^3+\Omega_{\Lambda}]$. As given in Eq.~\ref{eq:dv}, the volume-averaged distance $D_V(z)$ depends on both the Hubble expansion rate $H(z)$ and the angular diameter distance $D_A(z)$, allowing us to constrain the parameter pair $(\Omega_{\rm m},H_0)$. We adopt broad uniform priors of $\Omega_{\rm m}\sim\mathcal{U}(0.1,0.8)$ and $H_0\sim\mathcal{U}(20,100)\,\mathrm{km\,s^{-1}\,Mpc^{-1}}$.

\begin{figure*}
    \centering
    \includegraphics[width=0.45\textwidth]{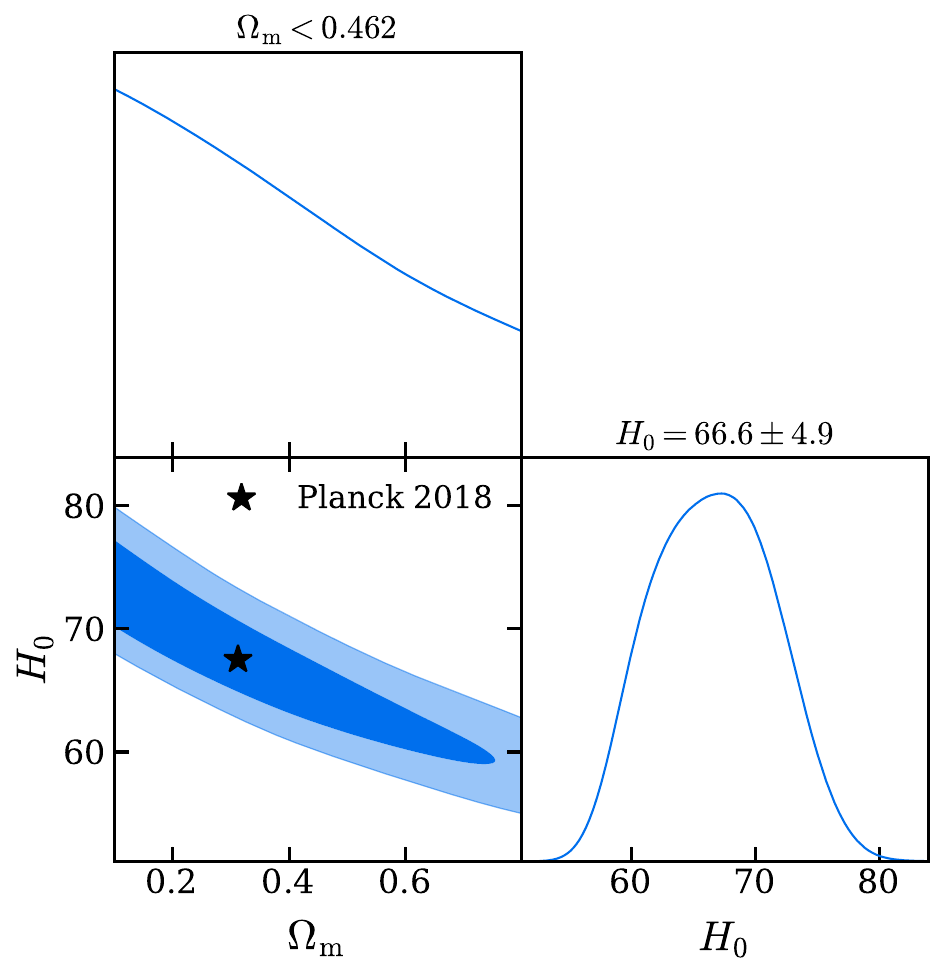}
    \hfill
    \includegraphics[width=0.45\textwidth]{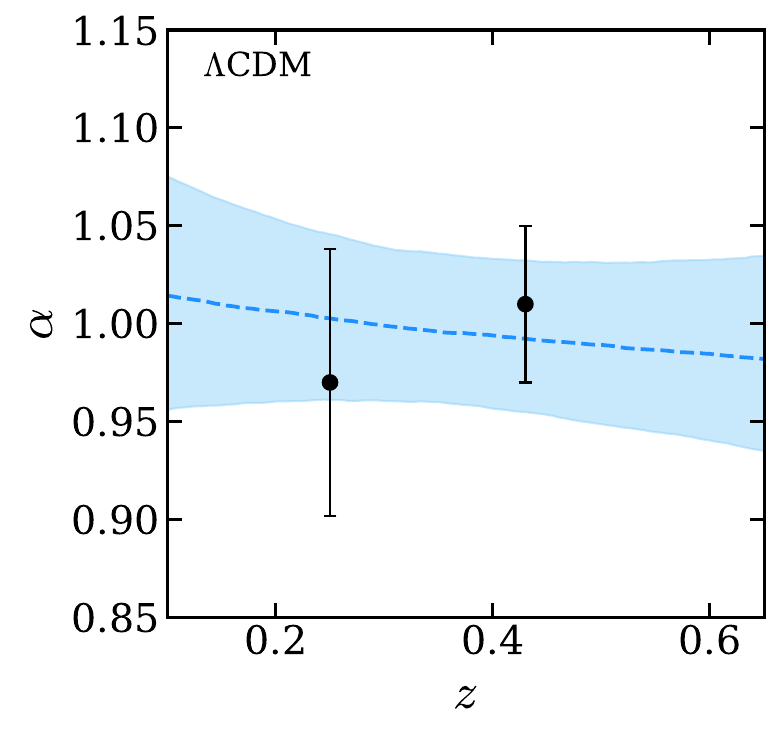}
    \caption{
    Constraints on the flat $\Lambda$CDM cosmological model from cluster isotropic BAO measurements.
    Left: Posterior constraints on $\Omega_{\rm m}$ and $H_0$, with filled contours indicating the $1\sigma$ and $2\sigma$ credible regions. The black star marks the Planck 2018 best-fit $\Lambda$CDM parameters.
    Right: Redshift dependence of the BAO dilation parameter $\alpha$ inferred from the posterior distributions. The dashed curve shows the best-fit prediction, while the shaded region indicates the 68\% credible region. The black data points represent the cluster BAO measurements at $z_{\rm eff}=0.25$ and $0.43$. The parameter $\alpha$ is defined relative to the fiducial Planck 2018 $\Lambda$CDM cosmology.
    }
    \label{fig:lcdm}
\end{figure*}

The resulting marginalized posterior distributions and the $1\sigma$ and $2\sigma$ credible regions are shown in Fig.~\ref{fig:lcdm}. Using the Planck-calibrated cluster BAO distances, we obtain $H_0=66.6 \pm 4.9 \,\mathrm{km\,s^{-1}\,Mpc^{-1}}$, while the matter-density parameter remains weakly constrained, with an upper limit of $\Omega_{\rm m}<0.462$ at the $68\%$ credible level. The Planck 2018 best-fit $\Lambda$CDM parameters lie within the region allowed by our cluster BAO measurements.
Despite relying on only two redshift bins, our analysis yields a substantially tighter constraint on the Hubble constant, with an uncertainty less than half that reported by Veropalumbo et al. \cite{veropalumbo2016measuring}, who obtained $H_0=72\pm13~\mathrm{km\,s^{-1}\,Mpc^{-1}}$ using 132,683 galaxy clusters from the WHL12 catalogue \cite{wen2012catalog}.

In contrast, the sensitivity to $\Omega_{\rm m}$ remains limited. Our upper limit of $\Omega_{\rm m}<0.462$ is consistent with the broader constraint of $\Omega_{\rm m}=0.32^{+0.21}_{-0.14}$ reported by Veropalumbo et al. \cite{veropalumbo2016measuring}. The relatively weak constraint on $\Omega_{\rm m}$ is mainly associated with the limited redshift coverage and uncertainties of our isotropic BAO distance measurements. The posterior distribution also exhibits a pronounced degeneracy between $\Omega_{\rm m}$ and $H_0$, with an orientation similar to that shown in Fig.~6 of Veropalumbo et al. \cite{veropalumbo2016measuring}.

We also examine the BAO dilation parameter $\alpha$ predicted by the inferred $\Lambda$CDM parameter distributions. Figure~\ref{fig:lcdm} shows its redshift dependence, together with the cluster BAO measurements at $z_{\rm eff}=0.25$ and $0.43$. The inferred values of $\alpha$ remain consistent with unity within the uncertainties over the redshift range probed by our cluster samples, indicating agreement between the measured BAO scale and the predictions of the fiducial Planck 2018 $\Lambda$CDM cosmology.

\subsection{Joint cosmological constraints}

As discussed in Section~\ref{sec:cosmo_constraints}, two cluster BAO distance measurements alone provide limited constraints on cosmological parameters. To investigate the implications of these measurements for dynamical dark energy, we perform a joint cosmological analysis by combining our uncalibrated BAO measurements, $D_V/r_d$, presented in Table~\ref{tab:alpha}, with CMB and SNe Ia observations. The CMB provides precise constraints on the physical baryon and cold dark matter densities, allowing the sound horizon $r_d$ to be determined within the joint analysis. SNe Ia serve as standardizable candles that probe the expansion history through the luminosity distance, $D_L(z)=(1+z)D_M(z)$, where $D_M(z)$ is the transverse comoving distance. Combining these complementary observations helps reduce parameter degeneracies and improve constraints on the dark energy equation of state \cite{desi2025dr2ii}.

We consider the Chevallier--Polarski--Linder (CPL) parametrization of dynamical dark energy, $w(z)=w_0+w_a z/(1+z)$, where $w_0$ is the present-day dark energy equation-of-state parameter, and $w_a$ characterizes its evolution with redshift.
At late times, neglecting radiation, the corresponding Hubble expansion rate in a spatially flat Universe can be expressed as
\begin{equation}
\frac{H^2(z)}{H_0^2} 
= \Omega_{\rm m}(1+z)^3
+
\Omega_{\rm DE}
(1+z)^{3(1+w_0+w_a)}
\exp\left(-\frac{3w_a z}{1+z}\right),
\end{equation}
where $\Omega_{\rm DE}=1-\Omega_{\rm m}$ is the present-day dark-energy density parameter. The standard $\Lambda$CDM model is recovered for $w_0=-1$ and $w_a=0$.

For the CMB, we use the Planck temperature (TT), polarization (EE), and temperature--polarization cross-correlation (TE) power spectra, adopting the Commander and SimAll likelihoods at low multipoles ($\ell < 30$) and the NPIPE CamSpec likelihood at high multipoles ($\ell \geq 30$) \cite{efstathiou2025power,rosenberg2022cmb}, together with the combined Planck and ACT DR6 CMB lensing likelihood \cite{madhavacheril2024atacama,maccrann2024atacama}. For the SNe Ia data, we adopt the Pantheon+ compilation of 1550 spectroscopically classified Type Ia supernovae \cite{brout2022pantheon+}. We perform a joint cosmological analysis combining the cluster BAO, CMB, and SNe Ia datasets using the corresponding likelihoods implemented in Cobaya \cite{torrado2019cobaya,torrado2021cobaya}.
We adopt uniform priors of $w_0\sim\mathcal{U}(-3,0)$ and $w_a\sim\mathcal{U}(-3,3)$ for the CPL dark energy parameters. We adopt broad uniform priors for the six standard cosmological parameters, $\{\Omega_bh^2,\Omega_ch^2,100\theta_{\rm MC},\tau_{\rm reio},n_s,\ln(10^{10}A_s)\}$, and fix the sum of neutrino masses to $\sum m_\nu=0.06\,\mathrm{eV}$.

\begin{figure}
    \centering    \includegraphics[width=0.7\linewidth]{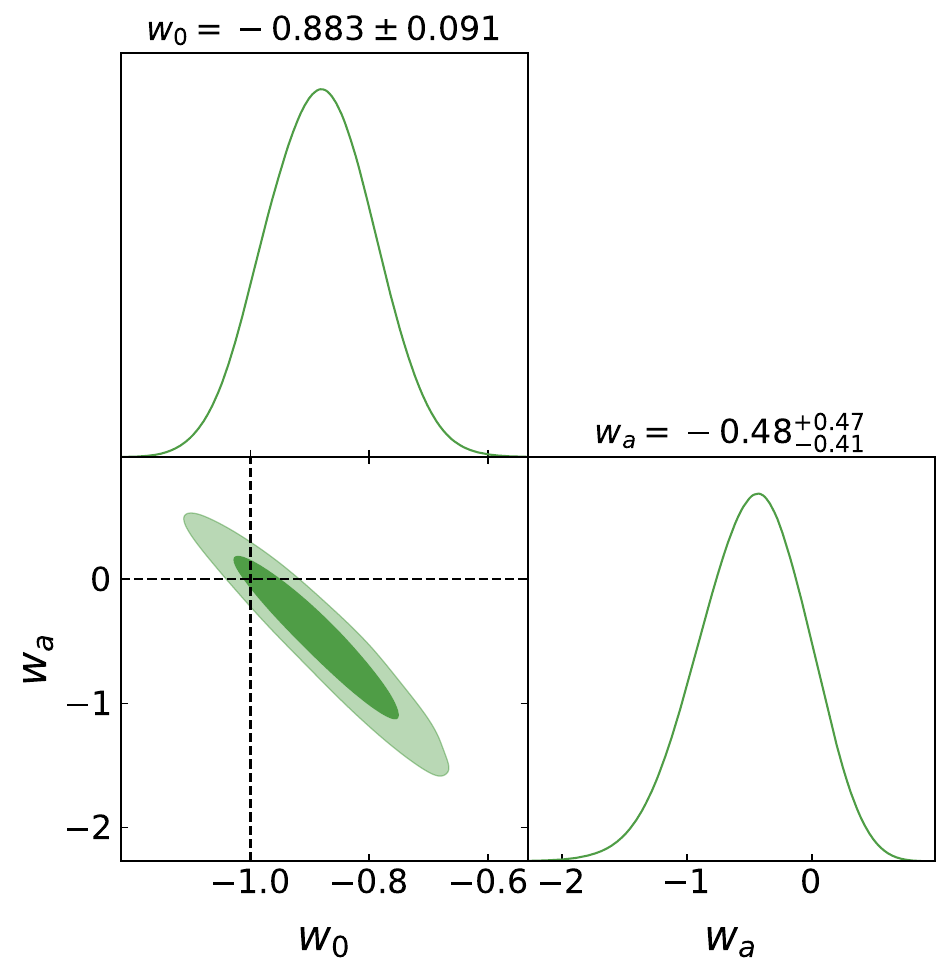}
    \caption{Joint constraints on the CPL dark energy equation-of-state parameters $w_0$ and $w_a$ from the combination of cluster BAO, CMB, and Pantheon+ SNe Ia data. The dark and light green contours represent the $68\%$ and $95\%$ credible regions, respectively. The dashed lines indicate the standard $\Lambda$CDM values, $w_0=-1$ and $w_a=0$.}
    \label{fig:cpl_com}
\end{figure}

Figure~\ref{fig:cpl_com} shows the joint constraints on the dark energy equation-of-state parameters $w_0$ and $w_a$ in the CPL model. From this analysis, we obtain $w_0=-0.883\pm0.091$ and $w_a=-0.48^{+0.47}_{-0.41}$ at 1$\sigma$ level. The standard $\Lambda$CDM model, corresponding to $w_0=-1$ and $w_a=0$, lies within the $68\%$ credible region, indicating that the joint constraints remain consistent with a cosmological constant.

\section{Conclusions}
\label{sec:conclusions}

In this work, we have measured the 2PCF of 169,679 galaxy clusters selected from the WH24 catalogue
\cite{wen2024catalog}, based on the DESI Legacy Imaging Surveys DR9 and DR10 \cite{dey2019overview}, constituting the largest spectroscopic galaxy-cluster sample to date. We divided the sample into two redshift bins,
$0.17 \leq z < 0.35$ and $0.35 \leq z < 0.5$, with effective
redshifts of $z_{\rm eff} = 0.25$ and $0.43$, respectively.
We also obtain linear bias
factors of $b = 2.51 \pm 0.03$ and $b = 2.77 \pm 0.03$ at
$z_{\rm eff} = 0.25$ and $0.43$, respectively. These values are
higher than those typically measured for galaxy samples,
consistent with the expectation that galaxy clusters preferentially
trace the highest-density peaks of the underlying matter distribution.

By fitting the BAO feature in the measured correlation functions,
we obtain dilation parameters of $\alpha = 0.97 \pm 0.07$
and $\alpha = 1.01 \pm 0.04$ at $z_{\rm eff} = 0.25$ and $0.43$,
respectively, corresponding to uncalibrated volume-averaged
distance measurements of $D_V/r_d = 6.70 \pm 0.48$
and $11.25 \pm 0.47$. These measurements are consistent with
previous cluster-based BAO results, DESI DR2 measurements at
comparable redshifts, and the distance--redshift relation
predicted by the standard $\Lambda$CDM cosmology.

Adopting the Planck 2018 sound horizon calibration,
$r_d = 147.09 \pm 0.26$ Mpc~\cite{aghanim2021erratum}, inferred within the
baseline $\Lambda$CDM model, we derive
$D_V(0.25) = 984.81 \pm 70.43$ Mpc and
$D_V(0.43) = 1654.91 \pm 68.64$ Mpc.
Using these calibrated distances, we obtain
$H_0 = 66.6 \pm 4.9$ km s$^{-1}$ Mpc$^{-1}$ within the
flat $\Lambda$CDM framework, in agreement with the Planck 2018
determination. However, the matter-density parameter remains
weakly constrained, with an upper limit of $\Omega_m < 0.462$
at the 68\% credible level.

We further perform a joint cosmological analysis by combining
our uncalibrated BAO measurements, $D_V/r_d$, with CMB and
Pantheon+ Type Ia supernova observations. This combination
provides tighter constraints on the cosmological parameters,
particularly those describing dark-energy evolution, by
reducing the degeneracies present in the cluster BAO measurements
alone. The resulting constraints remain consistent with
$\Lambda$CDM, which lies within the 68\% credible region in
the CPL parameter space, indicating that our measurements
do not require a departure from a cosmological constant.

Our results demonstrate the potential of galaxy clusters as
complementary tracers for BAO measurements. Despite the relative
paucity of galaxy clusters compared with galaxies, their higher
clustering bias enhances the clustering signal, while their
lower sensitivity to random motions can reduce the impact of
non-linear redshift-space distortions on BAO measurements~\cite{marulli2017redshift,veropalumbo2016measuring,veropalumbo2016clustering}. These properties make galaxy clusters
valuable tracers of large-scale structure, providing a
complementary cosmological probe to conventional galaxy-based
BAO analyses.

The potential of cluster-based BAO measurements will be further
enhanced by increasingly large cluster catalogues from wide-area
imaging surveys, such as Euclid \cite{collaboration2024mellier}, the Vera C. Rubin
Observatory's Legacy Survey of Space and Time (LSST) \cite{ivezic2019lsst},
the Nancy Grace Roman Space Telescope \cite{eifler2021roman}, the
Javalambre Physics of the Accelerating Universe Astrophysical
Survey (J-PAS) \cite{ascaso2016accurate}, and the China Space Station
Telescope (CSST) \cite{gong2023csst}. Combined with spectroscopic
observations from DESI \cite{desi2022overview} and proposed next-generation
facilities such as MegaMapper, the Maunakea Spectroscopic
Explorer (MSE), and the MUltiplexed Survey Telescope (MUST)~\cite{schlegel2022spectroscopic,sheinis2023maunakea,zhang2023conceptual}, these data will offer new opportunities
to obtain precise spectroscopic redshifts for substantially
larger samples of galaxy clusters. This will enable increasingly
precise BAO distance measurements over a broader redshift range,
further establishing galaxy clusters as a valuable complementary
probe of the cosmic expansion history. In particular, these
measurements will provide additional constraints on the nature
and possible redshift evolution of dark energy, contributing
to more stringent tests of the standard cosmological model.

\begin{acknowledgments}
We acknowledge the earlier studies of cluster baryon acoustic
oscillations by Veropalumbo et al.
\cite{veropalumbo2014improved,veropalumbo2016measuring}
and Hong et al. \cite{hong2015detection}.
We thank Wen \& Han \cite{wen2024catalog}
for making the WH24 cluster catalogue publicly available.

XS acknowledges financial support by FAPERJ Doutorado Nota 10 fellowship. PSF acknowledges the support from the start-up funding of Zhejiang University, Zhejiang provincial top level research support program. CAPB acknowledges financial support from the CNPq grant 306630/2025-7. RSG thanks financial support from FAPERJ grant No. 260003/005977/2024 - APQ1. JA is supported by CNPq grants (No. 307683/2022-2 and 448158/2025-6) and Funda\c{c}\~ao de Amparo \`a Pesquisa do Estado do Rio de Janeiro (FAPERJ) grant No. 299312 (2023). We also acknowledge the use of Cobaya, CAMB, Corrfunc, Astropy, CosmoBolognaLib,  GetDist, NumPy, SciPy, Matplotlib, IDL, and Mangle. The development of this work was aided by the National Observatory Data Center (CPDON).

\end{acknowledgments}

\printbibliography

\appendix

\end{document}